# Drastic changes before the 2011 Tohoku earthquake, revealed by exploratory data analysis

Tomokazu Konishi[1]

[1] Graduate School of Bioresource Sciences, Akita Prefectural University, Akita, Akita, Japan

Corresponding Author:
Tomokazu Konishi
Shimoshinjyo Nakano Kaidobata Nishi 241-538, Akita, 010-0195, Japan
Email address: konishi@akita-pu.ac.jp

## Abstract

**Background.** Japan is located in a subduction zone of several tectonic plates. Hence, frequent large earthquakes occur, and they are sometimes accompanied by tsunamis or eruptions. Because the damage caused by these disasters tends to be severe, there is an urgent need to predict large earthquakes in advance. Unfortunately, although much effort has been made with various types of measurements, a consistently accurate prediction system has not yet been realised. However, since there are few examples of statistical approaches to earthquakes, there seems to be a possibility of a breakthrough. Here, the possibility is examined.

**Methods.** I obtained the time of occurrence and magnitude of each recorded earthquake from open data reported every month by the Japan Meteorological Agency (JMA) and analysed them using exploratory data analysis (EDA). EDA is a statistical approach that uses data characteristics as a foothold. This reveals the hidden structure of the data, obtains the maximum amount of information from the data, and develops parsimonious models. For this purpose, a number of methods are available to visualise the characteristics of the data, such as the distribution.

**Results.** Earthquake intervals strictly followed an exponential distribution, confirming that earthquakes are essentially random events. The magnitudes followed a normal distribution, suggesting that the energy of the earthquake was determined. Once the distribution was determined and the parameters were clarified, changes in the distribution were clearly detected; the parameters of each of the distributions changed before large earthquakes. First, the intervals became shorter. This is partly because there were swarm earthquakes, which may indicate that the asperities at the plate boundaries were disrupted, and hence, the energy stored around them was destabilised. Second, the scale of the magnitude became large. In fact, the parameters of the magnitude distribution varied greatly on a monthly basis, which is quite different from what has been believed for a long time. With these changes, the likelihood of large earthquakes has increased. This indicates that a plate is a heterogeneous object and the hardness at the boundary fluctuates. Third, the magnitude of the moving average fluctuated greatly and became improbably high before the 2011 Tohoku Earthquake. In each case, because the original distribution was known, it was possible to accurately estimate the probability of such an anomaly. This property can be used to evaluate every change, and evaluation is important for the decision-making process required for prediction. In the past, EDA was difficult to calculate, but with the current availability of statistical packages, researchers can now easily try it for themselves. If applied to a larger number of different measurements, more accurate predictions can be made.

## Introduction

Japan is located in a complex subduction zone where several plates collide; therefore, earthquakes are frequent and strong, and occasionally trigger eruptions or tsunamis (Maruyama 1994; Stern 2002; Bird 2003; Fujii & Kazuki 2008; Japan Meteorological Agency (JMA) 2023a). The damage is so massive that the prediction of major earthquakes is urgently needed. For example, the 2011 Tohoku earthquake left more than 22,000 people dead or missing (Cabinet Office Japan 2023). At an elementary school, children and teachers who evacuated to the school grounds were swept away by the tsunami, and no one went back home. A nuclear power plant lost external power supply and exploded, and its removal is still pending. A wide area is inaccessible owing to radioactive contamination. These tragedies could have been avoided if we had known them just a few hours ahead of time. However, despite many efforts to accurately measure various factors, earthquake prediction has not been realised (Mogi 1998; Hayakawa 2012; Stark 2022). One possible reason for this failure may be the paucity of statistical approaches. Statistics can accurately predict the probability of random events, and it is difficult to make predictions without taking advantage of this characteristic.

This situation could be solved using exploratory data analysis (EDA). This is a parametric statistical approach that extracts valid information from data based on the characteristics of the data, such as data distribution (Tukey 1977; Croarkin & Tobias 2012). It differs from classical statistics in that it examines the properties of data without any prior knowledge. This method helps to maximise insight, uncover hidden structures, and explain the properties of a data set, and then develop parsimonious models. Previously, calculations for this purpose were complicated. However, the widespread use of computational statistics packages has made it possible for researchers to adopt this approach themselves. For example, microarrays can simultaneously measure the expression of various genes. A relative value was obtained, but the data were log-normally distributed (Konishi 2004). This property was used for normalisation, making the data comparable across different experiments and procedures (Konishi et al. 2008). This is because the distribution is universal and can be used as a system of units. I also showed that this property is derived from the mechanism by which cells regulate mRNA levels; this mechanism was modelled by integrating biochemical knowledge in terms of thermodynamics (Konishi 2005).

I applied EDA to a dataset of earthquakes in Japan. The parameters studied changed before the major earthquakes, indicating larger earthquakes were more likely to occur. These unstable properties seem to be different from the believed picture of Earth. EDA should make it easier to detect and evaluate these changes.

## Materials & Methods

From pdf files published monthly by the Japan Meteorological Agency (JMA 2023b), I extracted and analysed the time and magnitude of each perceptible earthquake. Since this is only a small portion of the various data they are measuring, the number was limited; therefore, I treated them without separating them by region.

An exponential distribution was found for the data (Stuart & Ord 1994; Bohm & Zech 2010). This is a distribution whose only parameter is $\lambda$, which represents the rate of events occurring. The density is expressed as $f(x)=\lambda e^{-\lambda x}$ for $x \geq 0$. It is heavily skewed to the left, with larger values having smaller densities. The mode is zero, and both the mean and standard deviation (SD) are $1/\lambda$. It is known as a distribution that represents the time interval of random events; if earthquakes occur, the time interval will follow this distribution, but if the randomness conditions change for some reason, $\lambda$ will change, and the distribution will be mixed.

The second is the commonly found normal distribution (Stuart & Ord 1994; Bohm & Zech 2010). This is a distribution whose density is denoted by $f(x) = (2\pi\sigma^2)^{-1/2}\exp(-(x-\mu)^2/2\sigma^2)$, with two parameters, $\mu$ and $\sigma$, which represent the location and scale, respectively. The location is estimated by the mean, and the scale is estimated by the SD. The sums of independent and identically distributed random numbers are known to have this distribution. If the logarithmic values of $x$ have this characteristic, it is called a lognormal distribution.

*Robust Estimation*

When data contain outliers, the estimation of parameters, such as the mean or SD, is affected by those outliers. Several calculation methods avoid this, and they are called robust methods. For example, the median or trimmed mean is a robust method for finding the mean, and the median absolute deviation (MAD) is a method for finding the SD. In addition, R provides functions such as *line* and *lm* to robustly find a linear approximation line.

*Quantile-Quantile (QQ) plot*

Histograms are often used to check the distribution; however, for a more rigorous check, a QQ plot was used. The ideal values of the distributions are obtained by specifying the number of data $n$ and the parameters. Originally, the QQ plot compares the data of each quantile of a dataset and the ideal data, but here it compares the whole sorted data against the ideal data. If a dataset obeyed the distribution model, a straight relationship was formed. For the exponential distribution, ideal values with $\lambda = 1$ were used. The slope of the QQ plot shows $1/\lambda$ of the data. For the normal distribution, the ideal values of $\sigma = 1$ and $\mu = 0$ were used. The slope of the QQ plot shows $\sigma$, and the intercept shows $\mu$. The parameters were identified with a robust calculation using the line function of R.

*P-value, probabilities in the cumulative distribution, and quantiles.*
The p-values are often used in statistical tests, but here they are used to evaluate the values. It is the probability of obtaining a value that is more extreme than a certain value. The p-value is the probability of cumulative distribution for a smaller value, and 1 minus that for the opposite (Fig. S1). The smaller this value, the stronger the anomaly (less likely to be obtained by chance). If z-score = (data-$\mu$)/$\sigma$ = -2, then the cumulative probability is the same as the p-value pnorm(-2) = 0.023. Similarly, if the z-score is 2, then pnorm(2) = 0.977; hence, the p-value is 1-0.977 = 0.023. Incidentally, to restore the quantile value, qnorm(0.023) = -2 was calculated.

*Distribution of the average*
According to the central limit theorem, the means of any random number are expected to be normally distributed with $\sigma = SD/n^{1/2}$, where $n$ is the number of data points for a mean (Stuart & Ord 1994; Bohm & Zech 2010). Using these characteristics, we can estimate the level that a moving average reaches only once a year. For example, the number of perceptible earthquakes before the main earthquake in Fig. 3C was 1,246. In R's notation, the quantile is qnorm(1-1/1246) = 3.15. The average number of measurements during the five days was 14.3. The median magnitude was 3.38, and the MAD was 0.276. From this, the upper possible quantile is calculated as median + MAD/sqrt(14.3)*3.15 = 3.57. This is indicated by the green line in Fig. 3C.
Conversely, we can evaluate how rare those moving averages are by estimating the p-value. The moving average reached 4.0 on 12 Jan 2011 in Fig. 3C, at which time the z-score of the average was (4.0-median)/(MAD/sqrt(14.3)) = 8.5. This is too large to estimate the p-value; its R notation is 1 - pnorm(8.5) = 0.

The calculations were performed using the R software package (R Core Team 2023), which is a free software environment for statistical computing and graphics that is suitable for EDA. A set of scripts for the installation and calculation are provided in the supplement.

## Results
As expected, the time interval followed an exponential distribution (Fig. 1A); here the mean of the interval was 9.3 hours. However, $\lambda$ sometimes increases (Fig. 1B); it has extremely increased three times in the past 20 years, each time corresponding to a major earthquake that caused a sequence of aftershocks (Fig. 1C); the change occurred because these earthquakes are out of the randomness of normal circumstances. They decrease rapidly according to the power series (Utsu et al. 1995), as can be seen in this panel. In addition, this appears in the data even before the main earthquake occurs. With swarm earthquakes, a distortion of the straight-line relationship occurs at the origin of the QQ plot (panel B), which is more apparent in shorter time intervals (Fig. 1D, 1E). Alternatively, anomalies near the origin are easier to detect by taking the logarithm of both axes (Fig. 1F). In the log-axes plots, the relationship between the linear regression line and $\lambda$ is changed (see the Fig. 1 legend). In addition, taking logarithms slightly changed the range of the data being focused on; therefore, the parameter estimation was also slightly altered.

It should be noted that this change would have occurred even if we had excluded swarm earthquakes. Most of the data were on a straight line with a larger $\lambda$ (panels B, D, E, and F). This means that before the 2011 Tohoku earthquake, for example, the frequency of all earthquakes increased.

The time course is presented in Fig. 2. The blue line is the moving average up to five days prior to the measurement. The green line indicates a level that is $2\sigma$ lower than the mean. The mean value of the intervals decreased below the green line three times. In "1" of panel A (blue), a sequence of earthquakes occurred inland from the epicentre of the 2011 Tohoku earthquake. Here, many smaller earthquakes had been occurring for more than a month, and they became strong enough to be recorded in the open data (JMA 2023b). At the same time, swarm earthquakes in Niigata were also recorded; here an M5.9 earthquake occurred on March 12. At "2" is an M5 earthquake at a different location followed by a series of aftershocks, and "3" is a swarm earthquakes that occurred just before the 2011 Tohoku earthquake. Panel B shows the data for the year before the eruptions on Miyake Island in 2000. Five earthquake swarms were detected. "1" and "3" were at Oita, where the highest M4 earthquake occurred on 4/29, followed by a main earthquake of M5.6 on 25 Apr 2001; "2" was a volcanic tremor of Mountain Usu, which started two days before the eruption; "4" was swarm earthquakes that occurred before Miyake Island erupted; and "5" was aftershocks associated with the eruption. Incidentally, the average time of intervals at A was 3.7 hours, and that at B was 8.7, showing that frequency increased before the 2011 Tohoku earthquake.

The distribution of magnitude followed a normal distribution (Fig. 3A), with a few outliers; hence, the parameters were found to be robust. In some cases, two different slopes were observed (Fig. 3B). This occurs when two states exist simultaneously in Japan: a lower half with a smaller location and scale, and the opposite. In fact, the 2011 data for the Tohoku region showed only a larger location and scale (panel B, nested).

The distribution of magnitudes sometimes changes drastically. In silent Oct 1998, the scale was 0.59 (panel A). However, the scale before the 2011 Tohoku earthquake was more than twice as large (Panel B). This indicates that larger earthquakes are more likely to occur. Table 1 shows the p-value for each magnitude, that is, the probability of a larger magnitude for each earthquake occurrence, and the expected annual frequency estimated from the rate of the interval. State 1 is for October 1998, and state 2 is before the Tohoku earthquake; for state 1, M9 is improbable, but for state 2, it is expected to occur once every two years. In fact, the QQ plot of magnitude for this year showed that M9 was not a surprising outlier (panel B). This would be far higher if it were in panel A.

The magnitude of the moving average can be evaluated. For example, the number of perceptible earthquakes before the main earthquake in Fig. 3C was 1,246. From this number, we can estimate the quantile in which the mean may exceed it only once a year (Materials and Methods). This is the green line in panel C at 3.57. The value in panel D was lower at 3.16.

Remarkably, this was exceeded many times in 2010-11 (Fig. 3C). This phenomenon would not be expected normally. For comparison, the year before the eruption of the 2000 Miyake Island (Fig. 3D); we can see that before the eruption, the average hardly exceeded the green line, except for the coincidence of earthquakes that occurred in several different places. In panel C, many M4-6 class earthquakes occurred along the Tohoku coastline (light red), and these increased the average. Earthquakes were activated in this area during the year, and they were commented on several times (JMA 2023b). Incidentally, we can estimate how rare those moving averages are in the form of p-values (see Materials and Methods). The moving average reached 4.0 on 12 Jan 2011; the z-score of this value was 8.5, and the p-value was zero. This means that such a large outlier is usually impossible.

Of particular importance are the earthquakes that occurred in the days preceding the main earthquake (panel C), which occurred at the seafloor north of the epicentre of the main earthquake (Ide et al. 2011; Editorial Committee 2013; JMA 2023b). Forty-three earthquakes were recorded, with an average of M5.2, without attenuation. Surprisingly, the z-score was 14. This could be an artefact caused by the fact that swarm earthquakes are related to each other and not independent events. This strongly indicates that many asperities that support each other coincidingly trigger many other ruptures nearby. There is no doubt that this destabilises the elastic energy stored in the periphery.

The finding of a normal distribution of magnitude is not consistent with the long-held Gutenberg–Richter law (Gutenberg & Richter 1944), which states that the logarithm of the energy of an earthquake is proportional to the logarithm of the frequency of its occurrence. I applied this law to real data, Oct 2015-Mar 2016, which was the period before the Kumamoto earthquake, and relatively few major earthquakes occurred. To make the comparison of the models fair, the expression of this law was updated (Fig. S2). Fig. 4 shows two QQ plots, where panel A is the normal distribution and panel B is the Gutenberg-Richter law. It is clear that the normal distribution model covers a wider range of data.

Unfortunately, there is not always a clear indication of an earthquake. For example, there was only a small change in magnitude before the 2016 Kumamoto earthquake (Fig. 1C, 4A), and no anomaly was observed in the moving average of magnitude; it only caused swarm earthquakes just before (not shown). This is especially true for volcanic earthquakes, as is the case for Miyake Island and Mount Usu in 2000 (Fig. 3D), the 2016 Mount Aso that could correspond to the Kumamoto earthquake, and the Tokara Islands in 2021 (not shown).

## Discussion

Knowing the appropriate distribution model for the phenomenon means that we can estimate what will happen as long as the randomness persists and can detect changes in the randomness as soon as they occur. EDA is an ideal approach to address this phenomenon. In Tukey's time, calculations were very difficult, and his book shows that a lot of effort was put into how to make the calculations easy (Tukey 1977). Today, however, anyone can easily perform this using R (Croarkin & Tobias 2012; R Core Team 2023). Another advantage is that we know the p-value of the random event we are dealing with; therefore, when something strange occurs, we can estimate its p-value and expectation. Using this, we can objectively evaluate each event. This is important information for prediction because when issuing an evacuation order, we must consider the level of economic loss it will cause. To deal with this, it is necessary to understand how unusual the phenomenon that is currently occurring is.

Here, we used open data, which is only a small fraction of the total observed data. If more data were available, a mesh of every 10 km, for example, could be analysed, which would perhaps reveal a blank area where earthquakes are less likely to occur. Such blank areas could be pinned areas of the stick-slip phenomenon (Scholz 1998; Matsukawa 2009), or they could be well-slip areas. If it is the former, it should be a reservoir of elastic energy and could become the epicentre of the next major earthquake. In addition, this would also reduce the noise level significantly because the information from other regions is essentially noise. This makes the detection of anomalies more sensitive.

Swarm earthquakes can be identified by changes in the randomness of the intervals. Aftershocks usually weaken rapidly (Fig. 1C) (Utsu et al. 1995). Perhaps they occur when part of the energy released from the main earthquake is transiently stored and then dissipated. However, isolated swarm earthquakes do not decay in such a way. This is because asperities, which store energy separately but support each other, rupture successively. In this case, they may occur over a wide area; in fact, in the case of "3" in the 2011 Tohoku earthquake (Fig. 3C), the extent was approximately 50 km wide (Ide et al. 2011; Editorial Committee 2013; JMA 2023b). This indicates that the area of the plate moved, and the conditions changed. This would destabilise a much larger reservoir of energy. In the cases identified here, many isolated swarm earthquakes preceded the main earthquake within a year. In particular, when isolated swarm earthquakes have a large magnitude, it indicates the proximity of a large main earthquake. In the Tohoku earthquake, the chain of ruptures led to a main earthquake and a series of subsequent major ruptures over a length of 100 km or more, which caused widespread seafloor changes that caused a tsunami (Ide et al. 2011; JMA 2023b).

The magnitudes were normally distributed (Fig. 3A, 3B, and 4A). This distribution is related to how earthquake energy is stored. The energy released by earthquakes is only a small portion of the energy of plate movements that carry continents. In the plate stick-slip phenomenon (Scholz 1998; Matsukawa 2009), a pinning asperity (Lay & Kanamori 1980) receives a small part of the movement energy and stores it as elastic energy. The strength depends on several properties of the asperity, such as shape, size, velocity, temperature, and hardness. The time it takes for an asperity to rupture is exponentially distributed, which may also be related to the stiffness of the plate. If some of these factors interact multiplicatively to determine the energy, the result is lognormally distributed. Perhaps this is what we are seeing, as the magnitude is a logarithmic number of the energy with a base of $1000^{1/2}$ (JMA 2020); hence, this normality will provide an idea of the model for energy accumulation.

Both the distributions of the time interval and magnitude kept changing. This indicates that a plate is a heterogeneous object; when its hardness changes at the contact surface, the parameters are altered. In fact, $\sigma$ of magnitude and $\lambda$ of the interval were extremely high in the year before the 2011 Tohoku earthquake (Fig. 3B). Under these conditions, earthquakes of large magnitude are more likely to occur (Table 1). This can last for a long time (Fig. 1B, 3B).

Although the classical Gutenberg-Richter law (Gutenberg & Richter 1944) is still considered reliable (Fujii & Kazuki 2008; Stark 2022), it did not coincide with the perceptible earthquake data (Fig. 4). This law was devised when measurements less than M3.5 were not yet accurate. Confirmation of consistency with the data was primitive (Fig. S2). It was also unclear what the formula meant physically, so it did not improve our understanding of the phenomenon. These are unavoidable shortcomings when considering the historical context. Nevertheless, this law could still be appropriate only if the missing small earthquakes are as frequent as it estimates. This would be an unverifiable assumption at present; hence, it cannot be introduced to keep the law. The normality was firm in the perceptible data (Fig. 3A, 3 B, and 4A), and it worked well to handle the distribution of magnitude.

In any case, the historical context is no excuse for not validating the law for eight decades. This medieval-like episode is evidence of how this field lacks statistics. In particular, the perception that parameter $b$, the slope of the linear regression (Fig. S2A), is always constant in this law is inconsistent with the observed facts (Hirose et al. 2002) (Fig. 3AB, Table 1), and it has to be revised. As is clear from the examples given here, in practice, the expectations of major earthquakes are always changing, and these can be easily estimated by statistical approaches.

Here, three anomalies were detected before the 2011 Tohoku earthquake. First, the intervals became shorter (Fig. 1B). This was often accompanied by swarm earthquakes, but they were not the entire cause of shortening. Rather, a situation with shorter intervals could have triggered swarm earthquakes. Second, the magnitude of these parameters increased (Fig. 3B). The first and second changes expanded the expectation of large earthquakes (Table 1). Third, the moving average of the magnitude was increasing to an improbable degree (Fig. 3C). In particular, there were large swarm earthquakes just before the main earthquake. Unfortunately, not all of these three always appear in all earthquakes as precursors. This may be because these are the characteristics of earthquakes associated with plate subduction. Perhaps other measurements of different characteristics should be used to predict earthquakes caused by different mechanisms. However, at the very least, people should be alerted to isolated swarm earthquakes. If overlapping anomalies are observed simultaneously, the alert level should be raised even higher. However, the approach of this study to analyse past data may be somewhat unfair in terms of prediction. The credibility of the ideas revealed here should be judged by what researchers will predict through statistical approaches in the future.

## Conclusions

EDA has been effective in detecting and assessing random changes. Although here, it has been applied to very simple events, useful information was provided. Therefore, I recommend EDA for other data analyses. Today, anyone can easily perform this task; above all, the researchers themselves would be experts who know the background of the data best. Here, instructions and scripts from the installation are supplied for R beginners. I hope that these are useful.

## Acknowledgements
I would like to thank Editage (www.editage.com) for English language editing.

# Figures and a table

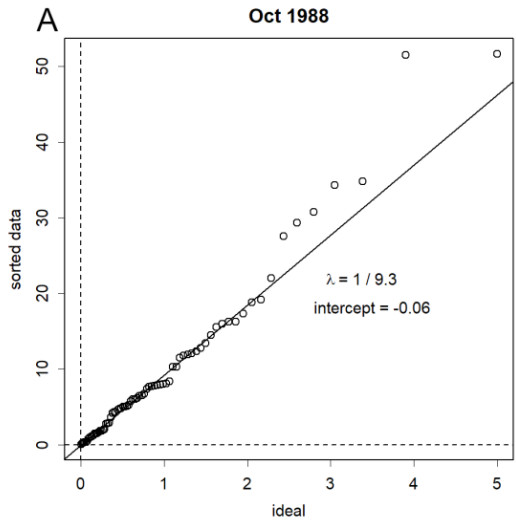

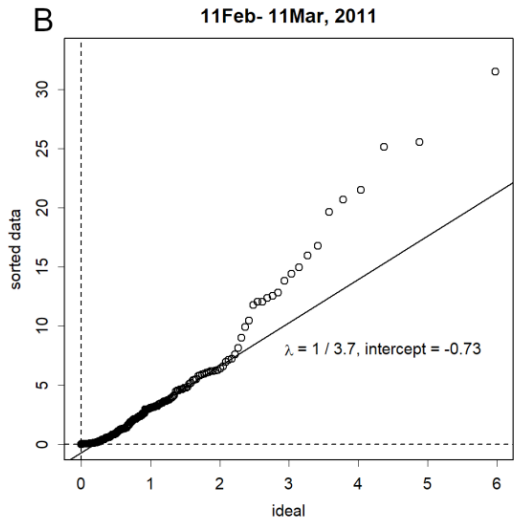

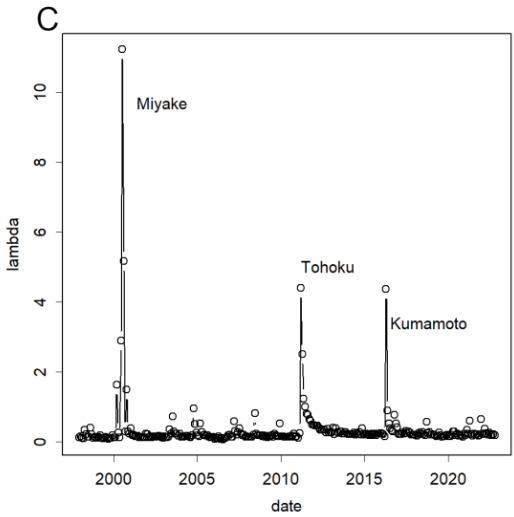

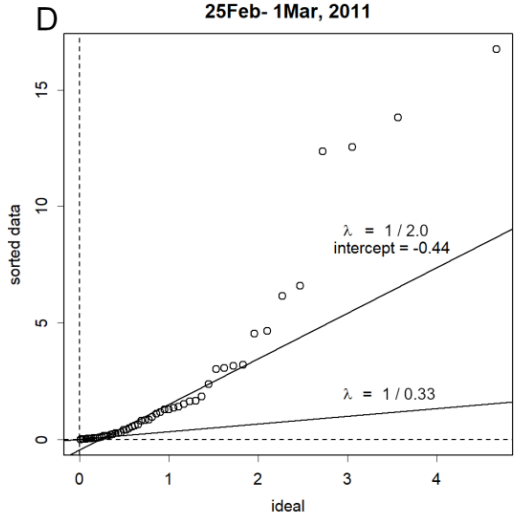

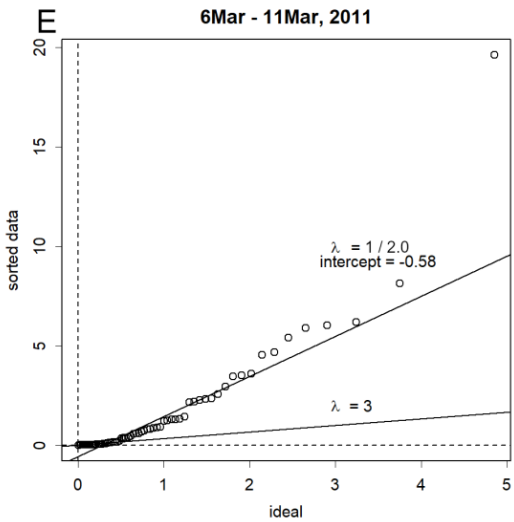

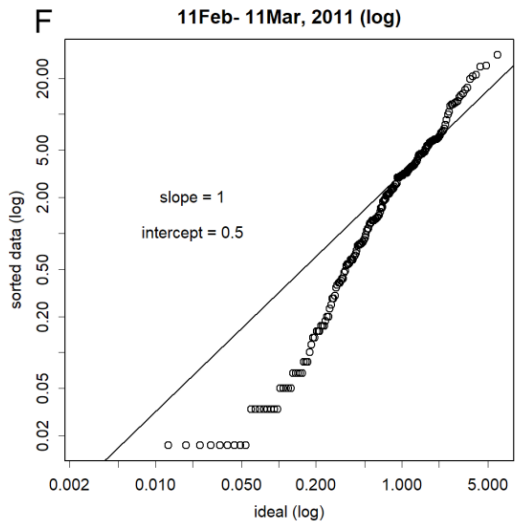

**Fig. 1** The time intervals (hour) of the earthquakes. (**A**) They strictly followed an exponential distribution. The reciprocal of λ is the mean and SD of the time interval. The parameter λ was not stable. (**B**) One month before the 2011 Tohoku earthquake. Not only λ but also the shape of the QQ plot changed around the origin. (**C**) λ sometimes changes drastically. The three peaks relate to the 2000 Miyake Island eruption, the 2011 Tohoku earthquake, and the 2016 Kumamoto earthquake. (**D, E**) The change became clearer when the period in the QQ plot is shortened. These are the distributions just before the main earthquake on 11 Mar. (**F**) One way to focus in on the anomaly near the origin is to take the logarithm (base = 10) of both axes. The data have a steeper slope when the high λ distribution is included. The slope of the linear approximation is 1 in principle. 1/λ would be 10^intercept.

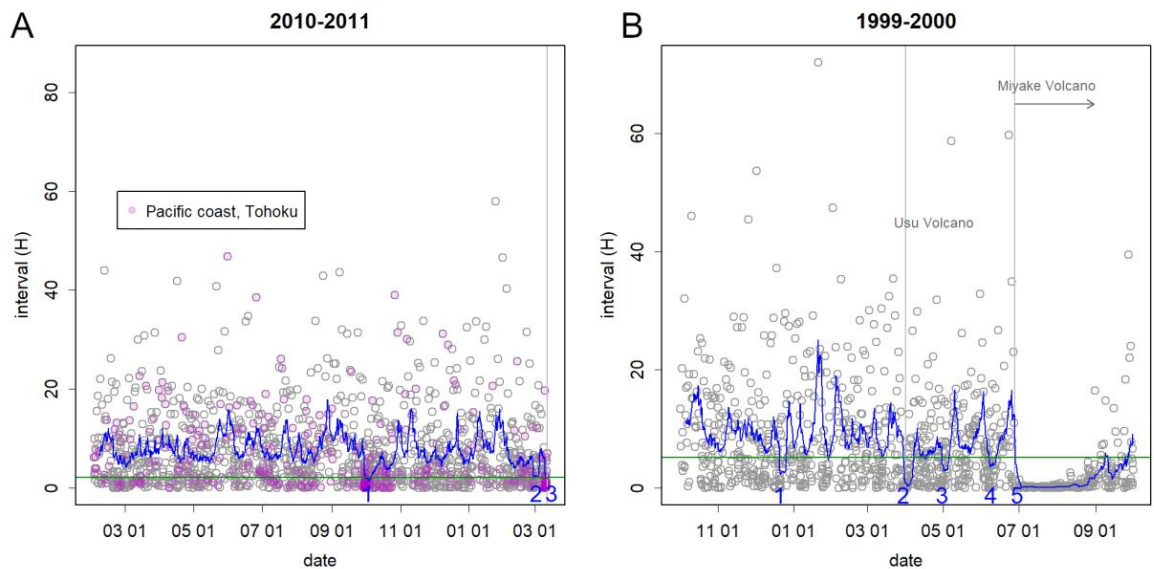

**Fig. 2** Changes in the time interval of earthquakes. (**A**) Before the 2011 Tohoku earthquake. The light red dots are Pacific Coastal areas of Tohoku. The blue line is the 5-day moving average. The green horizontal line is the level 2σ shorter than the mean of the time interval, estimated before the main earthquake. The moving average was lower than this level at the time indicated by 1 to 3 (blue). Records of small earthquakes after the main earthquake have not been published (JMA 2023b). (**B**) Levels before the Miyake Island eruption in 2000. An eruption of Mountain Usu in Hokkaido also occurred within this period (the left grey vertical line).

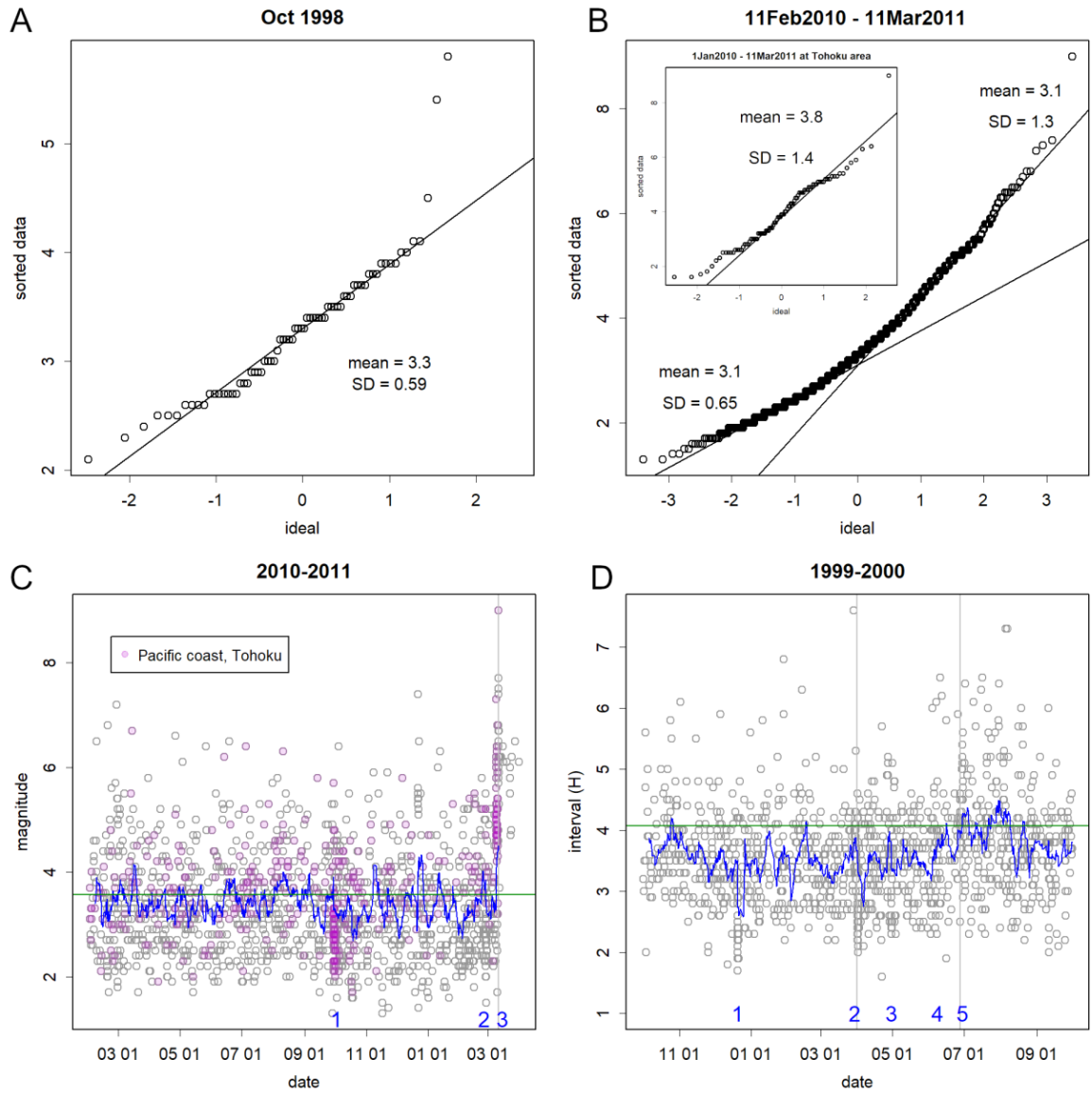

**Fig. 3** Distribution and time course of magnitude. (**A**) Distribution of magnitude, compared with the ideal distribution of the normal distribution. The parameters tend to change. (**B**) Before the Tohoku earthquake. In this case, the two states seem to be mixed. Nested is the Tohoku Pacific Coast from 1 Jan to 11 Mar 2011. (**C**) Time course before the 2011 Tohoku earthquake. Green is the level that the blue moving average is expected to reach only once a year. (**D**) Before the 2000 Miyake Island eruption.

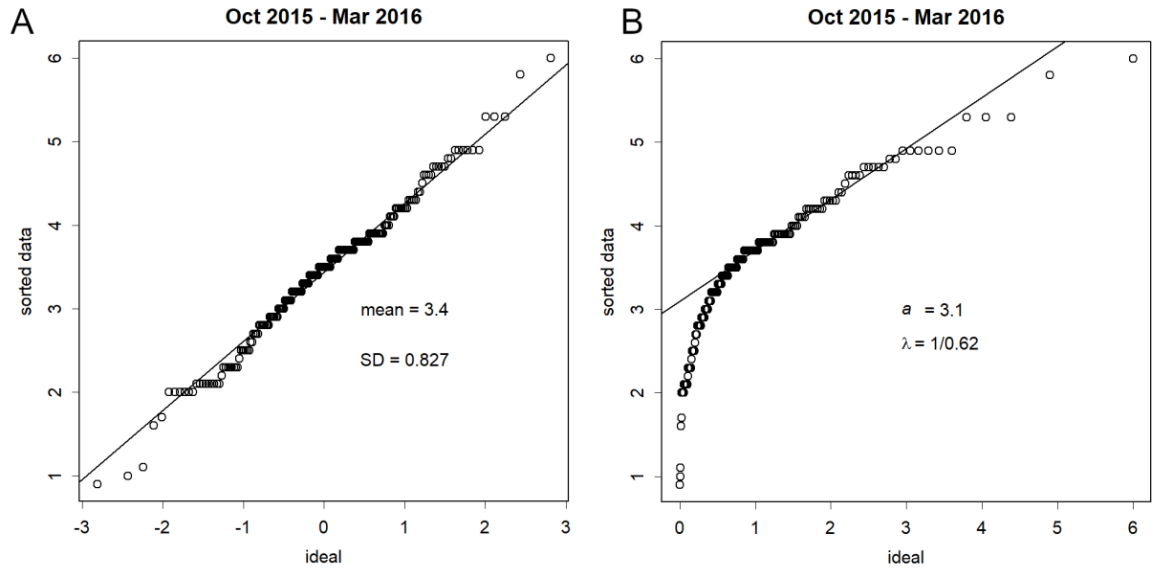

**Fig. 4** Comparison with Gutenberg-Richter law. (**A**) Normal QQ plot of magnitude, a linear relationship was obtained in the entire range of data. (**B**) Gutenberg-Richter law in an alternative presentation: QQ plot with the exponential distribution. The presentation is changed for comparison with panel A, see the legend in Fig S2 for details.

|   | state 1 rate = 1/9.3 (h$^{-1}$) location = 3.3 scale = 0.59 | | state 2 rate = 1/2.0 (h$^{-1}$) location = 3.8 scale = 1.4 | |
| --- | --- | --- | --- | --- |
| M | P-value | Expect/y | P-value | Expect/y |
| 3.5 | 0.37 | 236 | 0.58 | 621 |
| 4 | 0.12 | 90 | 0.44 | 590 |
| 4.5 | 0.021 | 18 | 0.31 | 494 |
| 5 | 1.9E-03 | 1.7 | 0.2 | 365 |
| 5.5 | 9.3E-05 | 0.085 | 0.11 | 238 |
| 6 | 2.2E-06 | 2.1E-03 | 0.058 | 136 |
| 6.5 | 2.7E-08 | 2.5E-05 | 2.7E-02 | 69 |
| 7 | 1.6E-10 | 1.5E-07 | 1.1E-02 | 31 |
| 7.5 | 4.7E-13 | 4.5E-10 | 4.1E-03 | 12 |
| 8 | 6.7E-16 | 6.3E-13 | 1.3E-03 | 4.2 |
| 8.5 | 0 | 0 | 3.9E-04 | 1.3 |
| 9 | 0 | 0 | 1.0E-04 | 0.45 |

**Table 1** P-values predicted from observed distributions and expected number of earthquakes per year.

## Supplementary Information

Besides Fig. S1 and S2, the compressed "data&script.zip" includes those four materials. This is available through figshare, doi: 10.6084/m9.figshare.22010279

*Rscript.txt*

    R code and instructions, including R installation.

*data2202.txt*

    Text-based data output from excel, Feb 2022.

*data2203.txt*

    Text-based data output from Excel, Mar 2022. Some characters are in the 2-byte code.

*.RData*

    Empty file to start R.
    Double-clicking this file will start R, specifying the working directory.

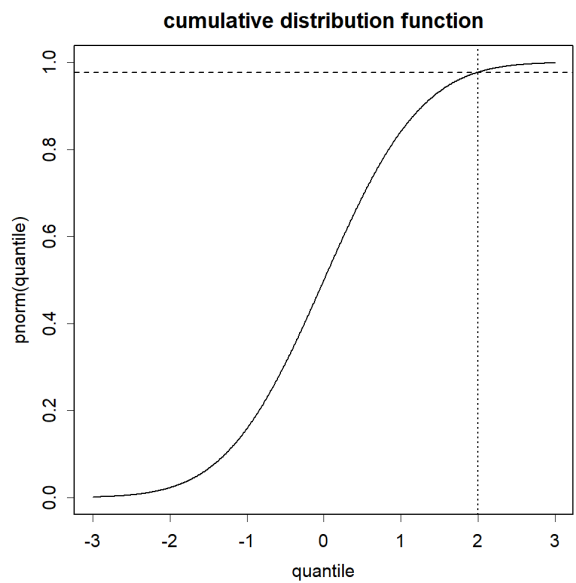

**Fig. S1** Cumulative distribution function of the normal distribution, $\mu = 0$, $\sigma = 1$. When the quantile is 2, the probability larger than this is 1-pnorm(2) = 0.023, which is the upper part above the dotted horizontal line.

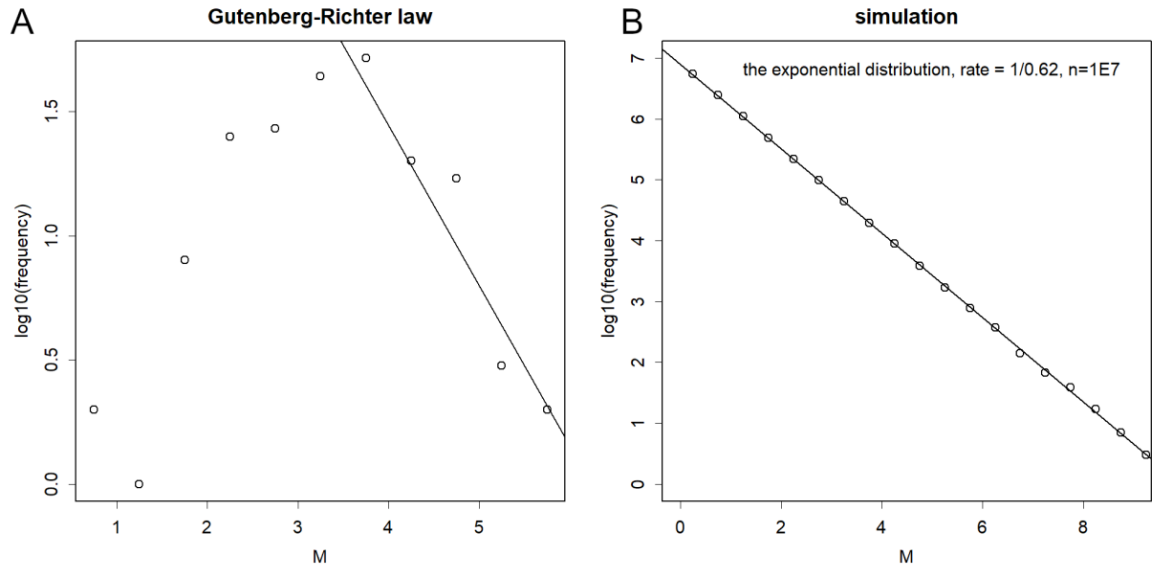

**Fig. S2** The Gutenberg-Richter law. (**A**) The relationship between logarithms of frequency and magnitude. The straight line is a linear approximation for the data over M3.5 ($a = 4.0$, $b = 0.65$). This is the original expression of Gutenberg-Richter law; however, this formula expression is less than ideal: as the data have to be classified by magnitude, the summarized data becomes so small that a precise verification is not possible. If we write this law as $n = 10^{(a - bM)}$, where $n$ is the frequency of each class, and $a$ and $b$ are constants, $n$ should be exponentially distributed with $\lambda = b/\log(e)$ when $a = 0$. In fact, exponentially distributed random numbers satisfy this law; (**B**) Data simulation, exponentially distributed random numbers according to the original Gutenberg-Richter law (the rate was deduced from Fig 4B). This would be a rough outline but would be sufficient to meet the necessary and sufficient requirements. Taking together, this law could be expressed as "the magnitude follows the exponential distribution with a shift at the origin". This was then used for Fig. 4B. The shift, $a$, may be caused by the amount of data that were not obtained, because zero magnitudes were defined by the sensitivity limit of the measurement; in principle, $a$ should be zero. Here the number of datapoints is 201, but if we use $n = 3.1E4$ for the ideal numbers and use the largest 201 for comparison, $a$ will be cancelled. Thus (3.1E4 – 201) would be the number of missing data with smaller magnitudes.